# Advanced pseudo-correlation radiometers for the Planck-LFI instrument

A. Mennella[(1)], M. Bersanelli[(2)], R.C. Butler[(3)], D. Maino[(2)], N. Mandolesi[(3)], G. Morgante[(3)], L. Valenziano[(3)], F. Villa[(3)], T. Gaier[(4)], M. Seiffert[(4)], S. Levin[(4)], C. Lawrence[(4)], P. Meinhold[(5)], P. Lubin[(5)], J. Tuovinen[(6)], J. Varis[(6)], T. Karttaavi[(6)], N. Hughes[(7)], P. Jukkala[(7)], P. Sjöman[(7)], P. Kangaslahti[(8)], N. Roddis[(9)], D., Kettle[(9)], F. Winder[(9)], E. Blackhurst[(9)], R. Davis[(9)], A. Wilkinson[(9)], C. Castelli[(10)], B. Aja[(11)], E. Artal[(11)], L. de la Fuente[(11)], A. Mediavilla[(11)], J.P. Pascual[(11)], J. Gallegos[(12)], E. Martinez-Gonzalez[(12)], P. de Paco[(13)], L. Pradell[(13)]

[(1)] *IASF-CNR, Sez di Milano*
*Via Bassini 15, 20133 Milano*

[(2)] *Univ. degli Studi di Milano, Dip. di Fisica*
*Via Celoria 16, 20133 Milano*

[(3)] *IASF-CNR, Sez di Bologna*
*Via Gobetti 101, 40129 Bologna*

[(4)] *Jet Propulsion Laboratory*
*4800 Oak Grove Drive, M/S 79-24, Pasadena, CA 91109-8099, USA*

[(5)] *University of Santa Barbara, University of California at Santa Barbara*
*Santa Barbara, CA 93106, USA*

[(6)] *MilliLab, VTT Information Technology*
*PO Box 1202, 02044 VTT, Finland*

[(7)] *Ylinen Electronics Oy*
*Teollisuuskatu 9 A, 02700 Kauniainen, Finland*

[(8)] *MilliLab, HUT Electronic Circuit Design Laboratory*
*PO Box 3000, 02015 HUT, Finland*

[(9)] *Jodrell Bank Observatory*
*Jodrell Bank, Macclesfield, Cheshire, SK119DL, UK*

[(10)] *University of Birmingham*
*Edgbaston, Birmingham, B15 2TT, UK*

[(11)] *Universidad de Cantabria, Departamento de Ingenieria de las Comunicaciones*
*Santander, E-39005, Spain*

[(12)] *Instituto de Fisica de Cantabria*
*Santander, E-39005, Spain*

[(13)] *Universidad Politecnica de Catalunya, E.T.S.E. Telecomunicacio Barcelona*
*Barcelona, E-08034, Spain*

**Abstract**

The LFI (Low Frequency Instrument) on board the ESA Planck satellite is constituted by an array of radiometric detectors actively cooled at 20 K in the 30-70 GHz frequency range in the focal plane of the Planck telescope. In this paper we present an overview of the LFI instrument, with a particular focus on the radiometer design. The adopted pseudo-correlation scheme uses a software balancing technique (with a tunable parameter called "gain modulation factor") which is effective in reducing the radiometer susceptibility to amplifier instabilities also in presence of small non-idealities in the radiometric chain components, provided that the gain modulation factor is estimated with an accuracy of the order of ± 0.2%. These results have been recently confirmed by experimental laboratory measurements conducted on the LFI prototype radiometers at 30, 70 and 100 GHz.

## 1. INTRODUCTION

Measurements of the CMB anisotropy has played a central role in observational cosmology during the past decade, following the discovery of the NASA satellite COBE which measured the CMB anisotropy over the whole sky with an angular resolution of ~ 7° and a signal to noise ratio of the order of 1 [1, 2]. After COBE many ground-based and balloon-borne experiments have dramatically increased detector performances thanks to the remarkable improvements in microwave and sub-millimetre detector technology, as well as in cryogenic technology (see e.g. [3] for a recent review). For these detectors the stability requirements are proportionally stringent calling for highly optimised instruments with very low susceptibilities to thermal and electrical parasitics which would contaminate the measured signal and the final CMB maps [4].

In coherent radiometric systems one of the major concerns is represented by $1/f$ fluctuations in the gain and noise temperature of the amplifier themselves, which degrade the receiver noise performances and introduce spurious correlations in the measured maps [5]. Such effects can be avoided if the post detection knee frequency, $f_k$, is significantly lower than the instrument scanning frequency, $f_s$, that in the case of Planck corresponds to the satellite spin frequency of ~ 17 mHz. If $f_k \gg f_s$ it is possible to mitigate such effects by applying destriping [6, 7] or map-making [8, 9] algorithms to the time ordered data provided that the level of $1/f$ noise is not too high. In the case of

Planck-LFI we have demonstrated that with a value of $f_k \leq 50 \text{ mHz}$ it is possible to maintain the increase in rms noise within few % of the white noise level and the excess power at low multipoles at a negligible level (about two order of magnitudes less than the CMB power) [6].

Differential receivers, such as the Dicke-switched scheme, greatly reduce the impact of amplifier instabilities by fast switching between the sky input port and a stable reference, sometimes given by another horn pointed at the sky. Dicke-type receivers have a long history in CMB observations and were successfully employed in the COBE-DMR instrument that first detected CMB anisotropies [1, 2]. Recently a new differential scheme (called pseudo-correlation radiometer) has improved over the classical Dicke scheme avoiding the presence of an active switch in the receiver front-end and increasing the sensitivity by a factor $\sqrt{2}$. In this design the sky signal, $T_{sky}$, is continuously compared to a stable reference load, $T_{ref}$, to reduce the effect of instabilities in the front-end amplifiers, while a fast (few kHz) switching between the two signals provides immunity from back-end fluctuations. Different versions of pseudo-correlation designs are being used for the second and third generation of space-based radiometric instruments for CMB anisotropy: NASA Wilkinson Microwave Anisotropy Probe (WMAP) and the Low Frequency Instrument (LFI) on board the ESA Planck mission.

In the case of the WMAP instrument, the radiometers directly measure temperature differences between sky signals from two widely separated regions of the sky [10, 11] while the LFI radiometers measure the difference between the sky and a stable internal cryogenic reference load cooled at about 4 K by the precooling stage of the High Frequency instrument. The offset of ~ 3 K between the sky and reference load signals is compensated introducing a "gain modulation factor", $r$, which balances the output in the on-board data reduction phase [12].

In this paper we discuss the design of the Planck-LFI receivers with a particular focus the susceptibility to $1/f$ noise fluctuations and to other major expected instrumental systematic effects. By means of a first-order analytical approach model we predict knee frequencies in the differenced data stream of the order of few tens of mHz also in presence of small non idealities in the radiometric chain [13]. These predictions have been confirmed both by numerical simulations and by experimental data taken with the LFI prototype models. The gain modulation strategy used to remove the sky-load offset involves simple on-ground data analysis of the radiometer data collected in undifferenced form to calculate the optimal gain modulation factor $r$ to be applied in the sky-load difference. Different approaches to calculate $r$ are discussed considering the constraints imposed by the available Planck scientific telemetry.

## 2. RADIOMETER ARCHITECTURE

Fig. 1 shows the baseline design of the LFI radiometers in which each feed-horn connects to an OrthoMode Transducer (OMT) that separates the incoming radiation into two perpendicular linearly polarised components that propagate independently through two parallel radiometric chains which are composed by an actively cooled front-end (20 K) and a 300 K back-end linked by waveguides.

In each radiometer, the sky signal and the signal from a stable reference load at ~ 4 K are coupled to cryogenic low-noise High Electron Mobility Transistor (HEMT) amplifiers via a 180° hybrid. One of the two signals then runs through a switch that applies a phase shift which oscillates between 0 and 180° at a frequency of 4096 Hz. A second phase switch is present for symmetry on the second radiometer leg but does not introduce any phase shift in the propagating signal. The signals are then recombined by a second 180° hybrid, producing an output which is a sequence of signals alternating at twice the phase switch frequency.

In the back-end of each radiometer the RF signals are further amplified, filtered by a low-pass filter and then detected. After detection the sky and reference load signals are integrated, digitised and then differenced after multiplication of the reference load signal by a so-called "gain modulation factor", $r$, which has the function to make the sky-load difference as close as possible to zero.

Neglecting asymmetries in the two radiometer legs we can write the radiometer output power in the following form:

$$\begin{aligned} p_{out} &= a\, G_{tot}\, k\, \beta\, [(T_{sky}^{hyb} + T_n) - r\,(T_{ref}^{hyb} + T_n)] \\ T_{sky}^{hyb} &= T_{sky} / L_{sky} + \left(1 - L_{sky}^{-1}\right) T_{phys} \\ T_{ref}^{hyb} &= T_{ref} / L_{ref} + \left(1 - L_{ref}^{-1}\right) T_{phys} \end{aligned} \quad (1)$$

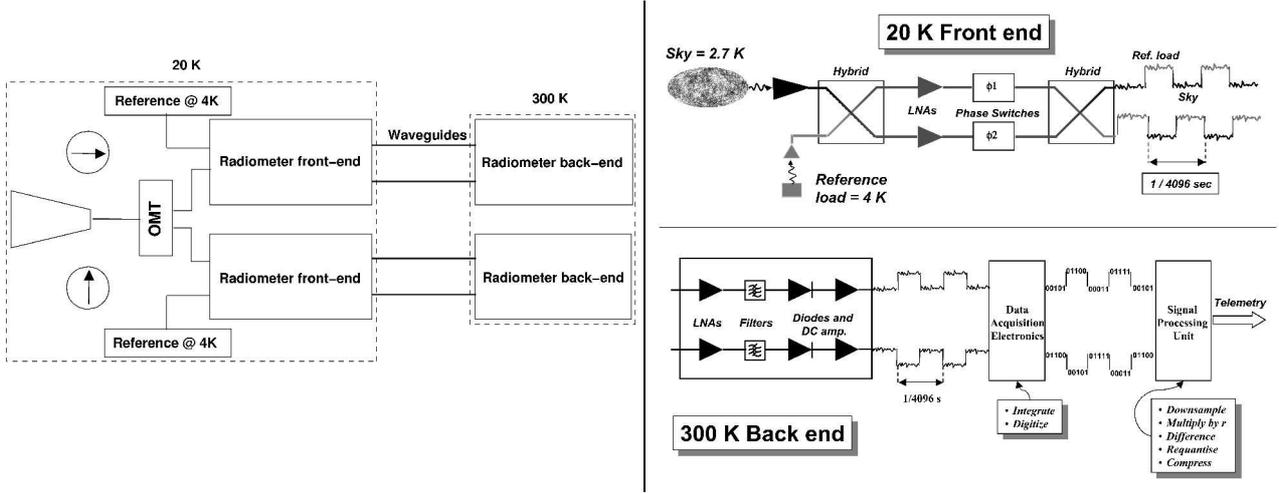

Fig. 1 - *Left*: schematic of the Planck LFI radiometer design. *Right*: detailed view of the front- and back-ends of the LFI receivers

where $T_{sky(ref)}^{hyb}$ represents the sky (reference) signal at the input of the first hybrid, $L_{sky(ref)}$ represent the insertion losses of the front-end passive components, $T_{phys}$ is the front-end physical temperature, $a$ is the detector proportionality constant, $G_{tot}$ the receiver total gain, $k$ the Boltzmann constant, $\beta$ the radiometer bandwidth, $T_n$ the radiometer noise temperature and $r$ the gain modulation factor.

From (1) it follows that $p_{out} = 0$ for $r = r_0^* = \dfrac{T_{sky}^{hyb} + T_n}{T_{ref}^{hyb} + T_n}$ ; averaging the radiometric output from each couple of detectors we have that for $r = r_0^*$ the radiometer sensitivity is given by:

$$\Delta T_{rms} = \sqrt{2} \, \frac{T_{sky} + T_n}{\sqrt{\beta \tau}} \tag{2}$$

which is independent of the level of the reference signal. Also in presence of slight unbalance in the radiometer properties it can be shown [13] that the dependence of the rms sensitivity to the level of the reference signal is at a level of $\partial \Delta T_{rms} / \partial T_{ref} < 10^{-5}$, which can be considered negligible.

## 3. SUSCEPTIBILITY TO 1/*F* NOISE FLUCTUATIONS AND OTHER SYSTEMATIC EFFECTS

Cryogenic HEMT amplifiers are known to have 1/*f* noise fluctuations in gain and noise temperature that can be described by $\dfrac{\Delta G(f)}{G} = \dfrac{C}{f^\alpha}$, $\dfrac{\Delta T_n(f)}{T_n} = \dfrac{A}{f^\alpha}$ (see, e.g., [13, 14, 15]), where $C$ is a normalisation factor, $A \sim C/(2\sqrt{N_s})$, $N_s$ represents the number of amplifier stages and $0.5 \leq \alpha \leq 1$. Following the approach described in [13] we can derive analytical formulas for the radiometer susceptibility to 1/*f* noise fluctuations in the radiometer gain and noise temperature, obtaining the following expression for the knee frequency, $f_k$:

$$\left\{ f_k = \frac{C}{T_{sky}^{hyb} + T_n} \sqrt{\frac{\beta}{2}} \left[ \left( T_{sky}^{hyb} + (1 + A/C) T_n \right) - r \left( T_{ref}^{hyb} + (1 + A/C) T_n \right) \right] \right\}^{1/\alpha} \tag{3}$$

which is zero for $r = r^* = \dfrac{T_{sky}^{hyb} + (1 + A/C) T_n}{T_{ref}^{hyb} + (1 + A/C) T_n}$; if $C/A \approx 2\sqrt{N_s} \gg 1$ then we have that $r^* \approx r_0^*$, which corresponds to the condition of null radiometric power output. In practice for $r = r_0^*$ gain fluctuations are cancelled and the residual 1/*f* noise is dominated by noise temperature fluctuations. In the left panel of Fig. 2 we show the expected knee frequency for the 30 and 70 GHz radiometers as a function of the noise temperature, $T_n$, and of the reference load temperature (indicated in the graph with $T_y$); the results clearly indicate the theoretical residual knee frequency is of the order of few mHz. The right panel shows the radiometer knee frequency as a function of the relative accuracy

$\delta r / r = (r - r_0^*)/r_0^*$ compared to the Planck-LFI design value of 50 mHz. The graph indicates that the necessary accuracy in the gain modulation factor calculation to meet the 1/$f$ noise requirements is driven by the highest frequency channel and is of the order of ± 2%.

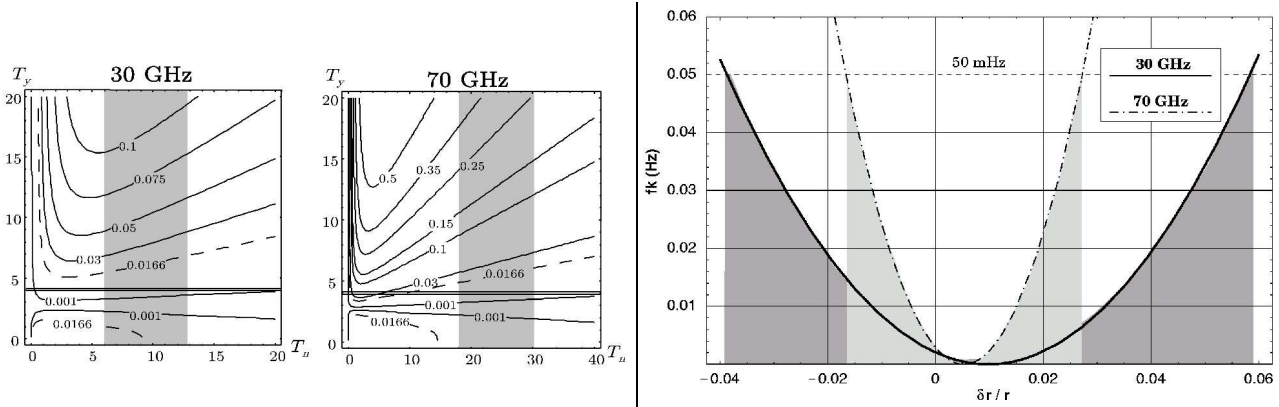

Fig. 2 - *Left*: Contours of equal $f_k$ (in Hz) as a function of the noise temperature ($T_n$) and of the reference thermodynamic temperature ($T_y$). The dashed contour refers to values for which $f_k = f_{spin}$ (the satellite spin frequency). The grey areas represent the range of typical noise temperature values and the double horizontal line represents the nominal reference load temperature. *Right*: knee frequency versus gain modulation factor relative accuracy.

Accurate analysis [13] of the effect of first order non-idealities in the radiometric chain components has shown that for a noise temperature match better than 10% and a gain match better than 1 dB the corrections to the zero-order knee frequencies shown in Fig. 2 are within ± 10%.

Apart from 1/$f$ noise, other instrumental systematic effects can be mitigated by balancing the offset in the radiometer output. In Planck-LFI, for example, systematic effects can be expected from thermal variations of the 20 K and 300 K stages and from input bias fluctuations of the front-end amplifiers. We calculate the susceptibility to various systematic effects following the approach described in [13]; any fluctuation in the terms appearing in (1) leads to a change in the observed signal which can mimic a 'true' sky fluctuation. If we denote with $\Delta T_{eq}$ the spurious signal fluctuation induced by a variation in a generic radiometer parameter $w$ we have that $\frac{\partial p}{\partial T_{sky}} \Delta T_{eq} = \frac{\partial p}{\partial w} \Delta w$, so that the susceptibility, $\phi_w$, can be written as $\phi_w = (\partial p / \partial w) \times (\partial p / \partial T_{sky})^{-1}$.

In Table 1 we report some characteristic values of the susceptibility of the LFI receivers to the major expected systematic effects. It is worth noting that the receiver susceptibility to fluctuations in the back-end temperature (as well as any other systematic effect at the level of the back-end stage) is virtually cancelled in the case $r = r_0^*$. Therefore the residual susceptibility depends on how accurately $r_0^*$ is calculated; the values indicated in Table 1 assume accuracies from ± 1% at 30 GHz to ± 0.2% at 70 GHz.

Table 1. Typical values of the susceptibility of LFI radiometers to the major expected systematic effects.

|  | 30 GHz | 44 GHz | 70 GHz |
|---|---|---|---|
| Susceptibility to front-end thermal fluctuations | 0.062 K/K | 0.046 K/K | 0.03 K/K |
| Susceptibility to back-end thermal fluctuations | ~ 0.003 K/K | | |
| Susceptibility to front-end bias fluctuations | ~ 0.009 K/mV | | |

Considering the electrical/thermal stability expected at the interfaces between the satellite and the instrument the values in Table 1 imply residual systematic effects on the final maps at a level < 1 µK per pixel peak-to-peak, which is compatible with the current LFI systematic error budget.

## 4. CALCULATION OF $r_0^*$ FROM MEASURED DATA

During a long-duration CMB measurement the gain modulation factor may need to be updated to compensate for slow variations in radiometer properties due to thermal/electrical drifts and ageing of the electronic components. If the offset balancing is done in software, then the value of $r_0^*$ can be recalculated using radiometer data acquired in "total power mode", i.e. before differencing is performed. In Planck-LFI the current telemetry bandwidth allows to download about 15 min of undifferenced data for each channel every day, in addition to the differenced data streams for all the channels[1]; therefore we have evaluated the feasibility to calculate $r_0^*$ with an accuracy of the order of ± 0.2% using a limited amount of total power data and the following three calculation methods: (i) $r_0^*$ calculated from the ratio of the average sky and reference load levels, (ii) $r_0^*$ calculated from the ratio of the sky and reference load standard deviations and (iii) $r_0^*$ calculated by minimising of the final differenced data stream knee frequency.

The various approaches have been analysed in detail considering the presence of a high level of 1/$f$ noise and also other systematic effects in the total power data (for further details see [12]).

In particular method (i) proves to be a simple and accurate way to calculate the gain modulation factor that is also highly insensitive from the presence of 1/$f$ noise and/or other systematic effects. The accuracy obtained with 15 minutes of data is much better than 0.01% also in presence of very high levels of 1/$f$ noise. With method (ii) the convergence is slower and depends on the presence of 1/$f$ fluctuations in the noise temperature. An estimate of the upper limit in the obtainable accuracy is given by $2/\sqrt{N}$, where $N$ represents the number of available data samples. The third method probes directly the 1/$f$ noise characteristics of the final differenced data, but can be limited in frequency resolution and accuracy when the available total power data is of limited length.

This gain modulation strategy is routinely applied with success in the Planck-LFI prototype radiometers. Here we present some results obtained with experimental data from the 30 GHz LFI *Elegant BreadBoard* (EBB) radiometer (see [16] for further details).

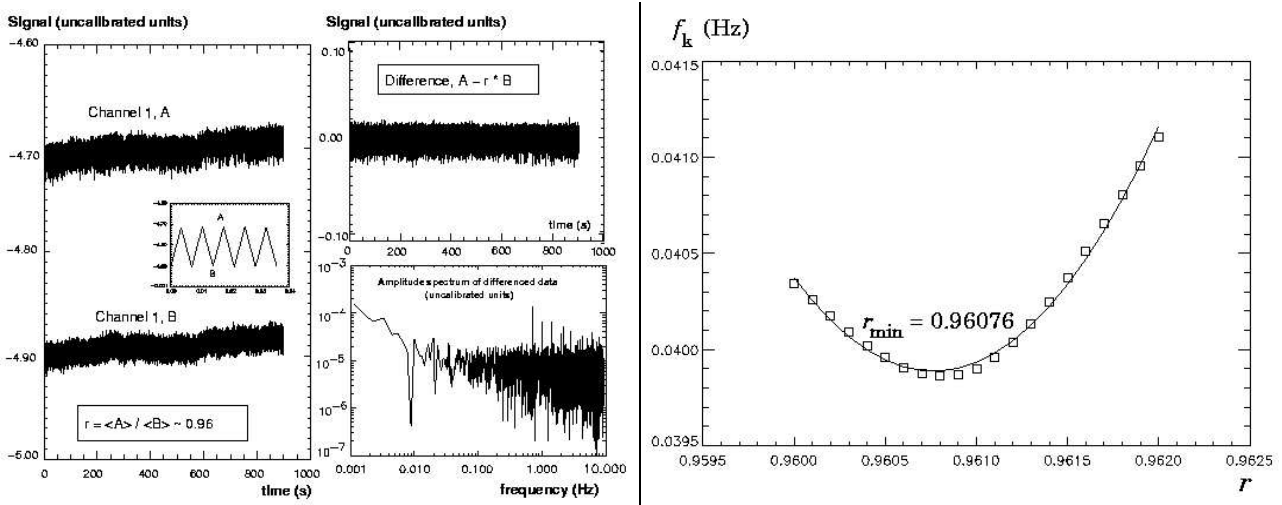

fig. 3 - *Left*: Measured data from the 30 GHz Planck-LFI prototype radiometer. The two panels on the right show the differenced noise stream in time and frequency domain. The final knee frequency is less than 50~mHz. *Right*: Behaviour of $f_k$ versus $r$ around the minimum.

---

1 An alternative data processing scheme is currently under study in order to download the complete total power data streams for all the channels

In the left panel of Fig. 3 we show the total power levels of one of the two output radiometer detectors, the differenced noise stream and its amplitude spectrum. The difference was performed applying a value $r = 0.96075$ calculated from the ratio of the average signal levels, so that a final knee frequency better than 50 mHz could been obtained.

We also calculated $r$ by taking the ratio of the standard deviation of the data streams, and by minimising $f_k$ (right panel in Fig. 3). In the first case we obtained $r = 0.95568$, while in the second case we found $r = 0.96076$ which is practically coincident with the one calculated with the ratio of signal levels. These results demonstrate how the pseudo-correlation scheme adopted for the LFI radiometers is effective in reducing the effect of amplifier instability to very low levels and that $r$ can be determined with very simple analysis of a limited portion of the total power radiometer data.

## 5. CONCLUSIONS

In this paper we have discussed the pseudo-correlation architecture adopted for the radiometers of the Planck-LFI instrument. With proper gain modulation the residual $1/f$ noise knee frequency is of the order of few mHz, provided that the gain modulation factor, $r$, is determined with an accuracy better than $\pm 0.2\%$. In this case also the susceptibility to other systematic effects (in particular those induced by fluctuation in the radiometer back-end) is greatly reduced.

The parameter $r$ can be calculated from the radiometer total power data using various calculation strategies; the most straightforward scheme uses the ratio of the average total power levels; this method proves quite simple, accurate and relatively immune from systematic effects like $1/f$ amplifier fluctuations, thermal effects etc. The application of this concept to prototype Planck-LFI radiometer data has shown that this radiometer scheme is effective and that the balancing condition can be calculated from a limited amount of raw radiometric data in total power mode.

Further studies will be aimed at a better understanding of the impact of systematic effects on the gain modulation factor accuracy and to continue the analysis of laboratory radiometer data in order to check our predictions. In addition we will continue with more realistic and detailed simulations.